\documentclass{article}

\usepackage[english]{babel}

\usepackage[a4paper,top=2cm,bottom=2cm,left=3cm,right=3cm,marginparwidth=1.75cm]{geometry}

\usepackage{amsmath}
\usepackage{amssymb}
\usepackage{amsfonts}
\usepackage{graphicx}
\usepackage[colorlinks=false, allcolors=blue]{hyperref}
\usepackage{algorithm}
\usepackage{algpseudocode}

\title{Mapping leadership and communities in EU-funded research through network analysis}
\author{Fabio Morea, Alberto Soraci - Area Science Park, Trieste, Italy \\
Domenico De Stefano, University of Trieste, Italy}

\begin{document}
\maketitle
\begin{abstract}

Horizon 2020 and Horizon Europe the EU programs supporting research and innovation through collaboration between companies, academic institutions, and research organizations. This paper introduces a novel methodology using open data on Horizon programs to analyse collaborations, leadership roles, and their evolution, with a focus on the North Adriatic Hydrogen Valley project in the hydrogen energy sector.
The methodology employs network analysis, transforming tabular data into weighted networks that represent collaborations between organisations. Centrality measures and community detection algorithms identify influential organizations and stable partnerships over time. To ensure robust and reliable results, the methodology addresses challenges such as input-ordering bias and result variability, while the exploration of the solution space enhances the accuracy of identified collaboration patterns.
  The case study reveals key leaders and stable communities within the hydrogen energy sector, providing valuable insights for policymakers and organizations fostering innovation through sustained collaborations. The proposed methodology effectively identifies influential organizations and tracks the stability of research collaborations. The insights gained are valuable for policymakers and organizations seeking to foster innovation through sustained partnerships. This approach can be extended to other sectors, offering a framework for understanding the impact of EU research funding on collaboration and leadership dynamics.

\end{abstract}

\section{Introduction}
\label{intro}
This paper introduces a novel methodology based on network analysis techniques for assessing the long-term impact of EU-funded research projects on companies, academic institutions, and research organisations. The application of network analysis tools to the innovation networks derived from collaboration to research project is not new and it has been used to study: topological properties of collaboration networks \cite{barber2006network}; the relation between network structure and the EU broad policy objectives \cite{breschi2006}; innovative dynamics at country-level \cite{balland}, \cite{MARUCCIA}; technological diversity \cite{Muscio}; \cite{cerqueti2023clustering} proposed the use of a rank-size approach for clustering complex networks derived from the collaboration in European project at organisational level. The novelty of the present application lies in its advanced data preparation using the \textit{European Science Vocabulary} \cite{euroscivoc} or \textit{EuroSciVoc} taxonomy, the incorporation of weighted networks, the application of community detection algorithms and temporal analysis to track collaboration over time. 

To demonstrate the practical application of the methodology, the paper includes a case study on hydrogen research and innovation, focusing on the development of hydrogen valleys and specifically the North Adriatic Hydrogen Valley project (NAHV).

Section 1 introduces the concept of hydrogen valleys and outlines the key research questions driving the study: Are there significant and stable communities within these networks? Which organisations assume leading roles? Do these leaders establish enduring communities, or do they primarily act as connectors between different groups? Addressing these questions requires a complex approach, employing network analysis and tackling challenges such as input-ordering bias and result variability, as detailed in Section 2 (Methodology).

Section 3 presents a case study focused on the development of hydrogen-related research and the current support to hydrogen valleys, using data from Horizon 2020 and Horizon Europe projects from 2015 to 2029. The results allow to identify organisation with leading roles, a community structure, and its evolution over time.  
 
\subsection{Hydrogen valleys}
\label{hvalleys}
A “Hydrogen Valley” is a geographical area where several hydrogen applications are combined into an integrated ecosystem that produces, exchanges, and consumes a significant amount of hydrogen, covering the entire hydrogen value chain: production, storage, distribution, and final use \cite{CleanHydrogen2021}. 

Currently more than 80 different hydrogen valleys are already ongoing or under development in Europe and in USA \cite{Weichenhain2021}. However, due to their relatively recent constitution \cite{Majka2023} few studied have been developed to map the hydrogen valley impact on a territory, and its development over the years. This small number of studies has been also assessed in 2023 by a Polish team \cite{Frankowska2023} by applying the bibliographic method and a quantitative analysis of the collected publications. The researchers identified 284 publications dealing with hydrogen valleys, with a systematic increase in the number of cited papers. Furthermore, from a deeper analysis, they found that the most frequently cited publications are studies presenting different types of technological solutions and concepts. This study indicated the existence of a significant research gap in the field of research on the creation and development of Hydrogen Valleys. Among the several, some other studies have been focused on the techno-economic assessment of green hydrogen valley \cite{Petrollese2022} providing multiple end-users and their LCA impact \cite{Concas2022}. For example, the South Africa Hydrogen Valley report \cite{southafrica2021hydrogenvalley} has examined the socioeconomic impact from the Hydrogen Valley project across multiple dimensions in terms of GDP, job market and tax revenue. 

The \textbf{North Adriatic Hydrogen Valley} or NAHV project \cite{NAHV_www} is one of the first transnational hydrogen valleys developed in Europe. It embraces the EU territories of Croatia, Slovenia and the Italian region Friuli Venezia Giulia and involves 37 partners based mainly on those countries. The project is co-financed by Horizon Europe programme and supported by the Clean Hydrogen Partnership \cite{HorizonJTI2022}. 

\begin{figure}[H]
    \centering
    \includegraphics[width=0.75\linewidth]{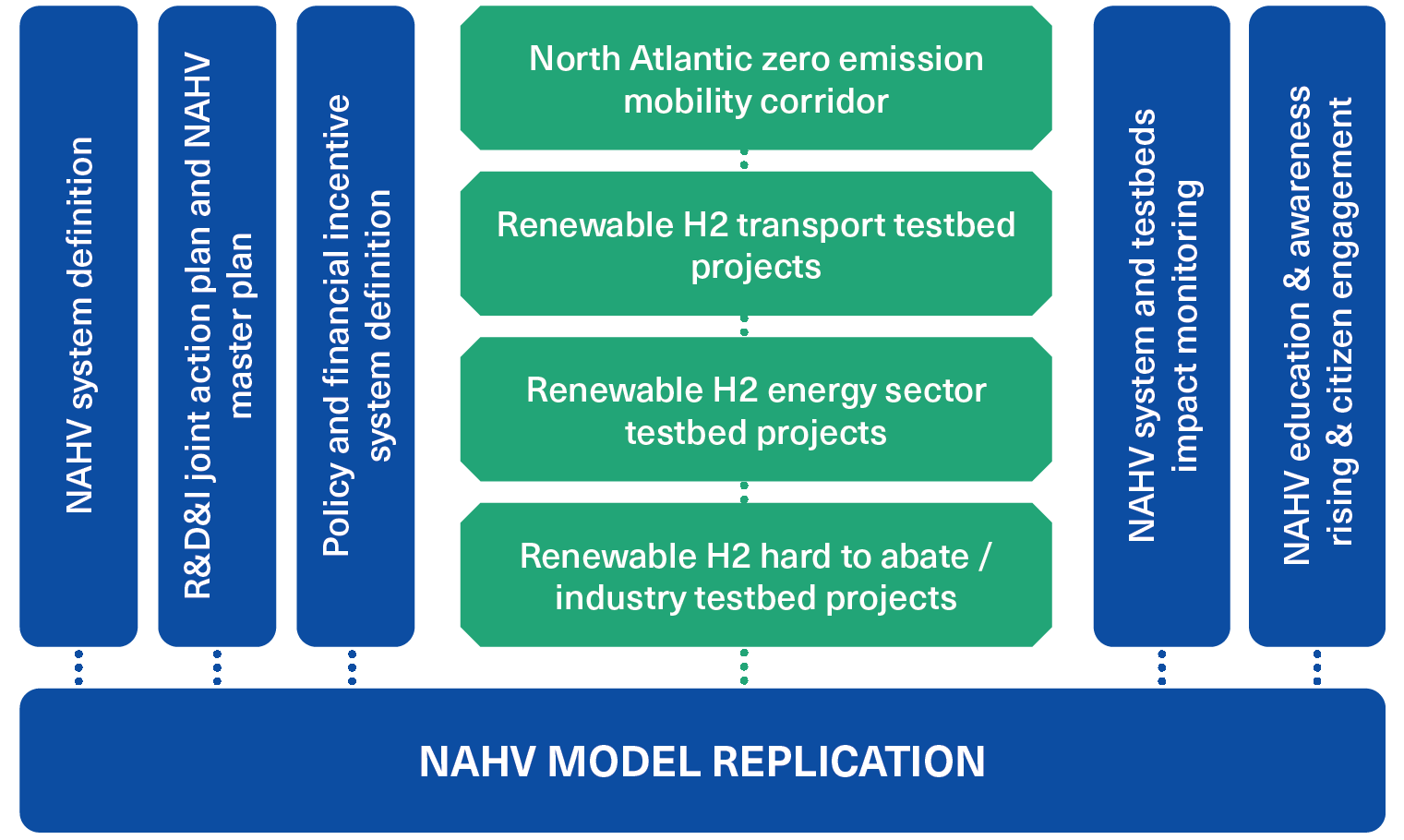}
    \caption{A schematic representation of the North Adriatic Hydrogen Valley project. Testbeds are primarily driven by industrial partners, while collaboration with policymakers, universities, and research organisations is crucial for enabling innovation. }
    \label{fig:NAHV_model}
\end{figure}

NAHV has been planned under the willingness to activate a long-lasting hydrogen economy. The heuristic structure of the NAHV is based on a systemic vision and the development of a holistic model of the valley. As showed in Figure \ref{fig:NAHV_model}, this heuristic approach considers not only innovation related to vertical “hard pillars” (which are typical of a hydrogen valley, such as production, transport, storage and distribution of hydrogen) but also to the "soft pillars“, which are regulatory framework, training and capacity building and involvement of civil society. Equally important is the level of cooperation among the various actors within the valley. NAHV aims also to develop and test a pathway of industrial transformation linked to green transitions \cite{Komninos2022} that will produce system innovation leading to a radical change of routines, and transform economic activities and industry ecosystems. 

\subsection{Collaboration-driven innovation in Hydrogen Valleys}
Hydrogen valleys are innovative ecosystems focused on hydrogen technology, leveraging regional and national initiatives to develop sustainable energy solutions. These ecosystems align well with the Quintuple Helix innovation model, which involves collaboration between five key actors: academia, industry, government, civil society, and the environment. By looking closer the hydrogen valley modus operandi, it is possible to observe that it is based on two main paradigms: (a) the open innovation model \cite{Chesbrough2003Open}, where the exchange of know-how among the several actors of the valley is stimulated and (b) the quintuple helix \cite{carayannis2010triple} model that is a model in which innovation is pushed by the integration in the classical triple helix model of the needs expressed by the civil society and media and by a targeted effort to solve a specific challenges \cite{casaramona2015toi}. In the context of Hydrogen valleys, the primary challenge is typically addressing climate change by driving the decarbonization of the economy.
According to this vision a trans-regional innovation ecosystems \cite{Hegyi2021, Pires2019} is needed to stimulate the interaction, creativity and innovation and to develop a strong and inclusive macro-regional ecosystems that support business growth, as well as the entrepreneurial environment necessary to foster more innovative regional economies \cite{Agrawal2014} and to increase competitiveness of small- and medium-sized enterprises (SMEs). To reach this wide and ambitious scope, it is necessary to put in action a combined strategy \cite{Soraci2009} based on:

  \begin{itemize}
      \item Developing Research, Development, and Innovation (R+D+I) within the Hydrogen Valley ecosystem to foster collaboration and identify synergies that can be effectively implemented. 
      \item Establishing a common development framework to enhance the business ecosystem’s capacity. This includes encouraging the creation of new businesses and attracting investment to address gaps in the hydrogen value chain, particularly by supporting new ventures \cite{fednor2024regional}.
  \end{itemize}

Innovation, in this context, is inherently driven by collaboration, particularly through partnerships between industry and academia, which combine diverse approaches and cover the entire Technology Readiness Level (TRL) scale. However, a quantitative understanding of collaborations and their evolution over time remains elusive. This paper seeks to address this issue by providing a methodology to identify and gauge collaboration, offering insights into the mechanisms that support innovation and the dynamics of innovation ecosystems.

From the perspective of individual organizations, understanding the dynamics of innovation at the regional or international level can be a good opportunity. Consider, for example, an organization involved in a new hydrogen valley project: it must not only focus on the development of its own project activities, but it may benefit from identifying partners and competitors that are engaged in other hydrogen valleys. This broader awareness may help  identifying synergies, risks, and market opportunities. 

Policymakers adopt yet another perspective. They have a strategic vision to align local, national, and European resources, aiming for medium- and long-term  impact of the innovation ecosystems. Achieving this requires a clear understanding of how collaborations emerge between industry and research, who the key stakeholders are, and how these collaborations evolve over time.

This paper addresses the elusiveness of innovation proposing a methodology that is suitable for two approaches: project-specific (e.g., examining the NAHV project to gain deeper insights into its role within the broader landscape of EU-funded hydrogen research) and a policy-oriented (e.g., analysing how Horizon grants are shaping competitive ecosystems in hydrogen sector). In both cases, two research questions are essential:
\begin{enumerate}
    \item Which organisations are the most influential in driving research and innovation? Identifying these key players — whether companies, academic institutions, or research organisations — is important for other organisation that want to be in contact with them in future projects. Moreover, policy makers may be interested in assessing whether the policy put in place in their territory has produced an improvement of leadership roles over the years.
    \item Are EU-funded projects fostering partnerships that extend beyond the duration and scope of individual projects? If such long-term communities exist, they serve as vital indicators of the open innovation and quintuple helix models effectiveness. Stable communities suggest robust exchanges of ideas and collaboration, which are essential for sustaining innovation and achieving the long-term goals of hydrogen valleys. 
\end{enumerate}

The methodology presented in Section \ref{met} aims to provide quantitative answers to the research questions above. Key requirements of this approach are that results must be independent of contingent factors (such as software implementation or ordering of the input data) and tested for validity. Additionally, when multiple algorithmic options are available, the selection must be data-driven and performance-oriented, ensuring that the chosen algorithm yields the most interpretable and reliable outcomes.
 
\section{Methods}
\label{met}
The proposed methodology builds on well-established network analysis techniques, such as centrality measures and community detection, while incorporating novel elements to enhance the stability and quality of results, and to produce a temporal analysis. Notably, a new method for exploring the \textit{solution space} is introduced, along with a comprehensive quality check process to select the most appropriate algorithm, validate the results, and manage input-ordering bias. When multiple solutions emerge, a \textit{consensus community detection} approach is employed to consolidate them into a single, coherent outcome.
 
The methodology foresees 6 main steps: data acquisition further discussed in section \ref{met_data_source}, data preparation and enrichment (section \ref{met_data_prep}), network analysis (section \ref{met_network}), centrality measures (section \ref{met-centrality} community detection (section \ref{met_com_det}), and temporal analysis of communities (section \ref{met_tempo}). 
An overview of the process is provided in Algorithm \ref{algo:overall}
\begin{algorithm}
\caption{Methodology for analysing research collaborations using CORDIS data}
\label{algo:overall}

\begin{algorithmic} [1]
\State \textbf{Data acquisition:} 
\State Access CORDIS and download all files related to Horizon 2020 projects 
\State Access CORDIS and download all files related to Horizon Europe projects
\State Merge corresponding tables for Organisations $\mathbf{O}^*$, Projects $\mathbf{P}^*$ and Topics $\mathbf{T}^*$ 
\State Enrich data with AI generated keywords and categories
\State Select relevant topics $\mathbf{T} \subset \mathbf{T}^*$
\State $\mathbf{P} \subset \mathbf{P}^*(\mathbf{T})$  \textit{subset projects by selected topics}
\State $\mathbf{O} \subset \mathbf{O}^*(\mathbf{P})$ \textit{subset organisations that are involved in selected projects}

\Statex
\State \textbf{Calculate weights matrix $\mathbf{W}$}
\State \textbf{Data preparation:} 
\State  $y_1 \gets$ min(\textit{project start year} $\forall p \in P$)
\State  $y_2 \gets$ max(\textit{project end  year} $\forall p \in P$)

\For{$y = y_1$ to $y_2$}
\For{\textit{each project } $p \in P$}
\State $f_{p,y} \gets ($ \textit{duration of} $p$ \textit{within year} $y ) / 365$
\For{\textit{each organisation $o$ involved in project} $p$}
\State $\mathbf{W}(o,p,y) \gets f_{p,y} * $ \textit{total cost of} $p$ 
\EndFor
\EndFor
\EndFor

\For{$y = y_1$ to $y_2$}
\State $\mathbf{A}_y \gets \mathbf{W}_y^T \mathbf{W}_y$ \textit{Calculate adjacency matrix for one-mode network}
\State $G_y \gets $ network nodes and edges from $\mathbf{A}_y$ 
\State Calculate centrality measures of $G_y$ and associate values to each organisation
\State Community detection $C_y \gets$ \textit{a valid and stable partition of} $G_y$
\EndFor

\Statex
\State  \textbf{Temporal Analysis}
\For {$y = y_1$ to ($y_2 - 1)$}
\For {$c_i \in C_{y1}$}
\For {$c_j \in C_{y1 +1}$}
\State assess correlation between $c_i$ and $c_j$ as \textit{continue, split, merge, mix, disjoint}
\EndFor
\EndFor
\EndFor
\State \textbf{Generate global community labels}
\State \textbf{Output:} community labels and centrality measures associated to each organisation $o \in \mathbf{O}$
\end{algorithmic}
\end{algorithm}
 
\subsection{Data acquisition}
\label{met_data_source}
The source of data for this study is CORDIS (Community Research and Development Information Service), the European Commission's primary public repository and portal to disseminate information on all EU-funded research projects and their outcomes \cite{CORDIS}. It provides open access to comprehensive data on projects, including objectives, participants, funding details, and results. For the purpose of this study, two datasets have been extracted: Horizon 2020 (projects starting from 2014 to 2020) \cite{Horizon2020} and Horizon Europe (projects starting from 2021 to 2027) \cite{HorizonEurope}. The two datasets have the same structure, composed of several tables, the most relevant being the organisations table denoted as $\mathbf{O}^*$ and the projects table denoted as $\mathbf{P}^*$. 

Each project is characterized by a unique identifier (\texttt{projID}), its acronym, title, start date, end date. Moreover, each project is associated with a long textual field, describing its objectives, and with structured categorical information on the call and funding scheme. A separated table associates each project with one or more topics, encoded according to the European Science Vocabulary (\textit{EuroSciVoc}), the multilingual taxonomy representing scientific fields developed by the EU’s Publications Office. 

There is a many-to-many relationship between organisations and projects (i.e. a project has several organisations, and an organisation can be part of several projects). In table  $\mathbf{O}^*$ each organisation is identified by a unique identifier (\texttt{orgID}), a name, detailed geolocation information. In addition, the table records the role of each organisation in each project (coordinator, participant, or associated partner), as well as the total costs incurred by the organisation within each project (\texttt{totalCost}) and the corresponding eligible contribution from the Horizon programme (\texttt{netEcContribution}). 

Data representation can be optimized by merging the corresponding tables for Horizon 2020 and Horizon Europe, to obtain the following core data structures: 

\begin{itemize}
    \item table $\mathbf{P}^*$ listing all projects, their \texttt{projID}, start and end dates, reference to the Horizon calls for proposal, keywords, and a long textual descriptions of project objectives;
    \item table $\mathbf{O}^*$ listing all organisations, their \texttt{orgID} location, their role in the project, and monetary values \texttt{totalCost} and \texttt{netEcContribution};
    \item table $\mathbf{T}^*$ associating each project in $P$ to one or more keywords of the \textit{EuroSciVoc} taxonomy. 
\end{itemize}

\subsection{Data preparation}
\label{met_data_prep}

The initial step in data preparation involves identifying the subset of projects $\mathbf{P} \subset \mathbf{P}^*$ on which we aim to focus, based on the topics from that are relevant for the objectives of the analysis. Consequently, the other tables can be filtered to ensure that only pertinent data is retained in the form of  $\mathbf{O} \in \mathbf{O}^*$ (organisation name and location). 
The core information for our analysis the weights table denoted as $\mathbf{W}$, which encodes the annual effort each organisation $o \in \mathbf{O}$ puts into each project $p \in \mathbf{P}$ in a given year. 
A proxy for the values in $W$ can be derived from either  \texttt{netEcContribution} or \texttt{totalCost}. The former indicates the total amount of public funding received by an organisation upon project completion and serves as a useful proxy for how much the Horizon grant promotes collaboration. The latter reflects the total cost incurred by the organisation to complete the project. For non-profit organisations involved in research projects, these values often coincide, as the Horizon grant may cover $100\%$ of the costs. However, for private companies investing in pilot projects, such as in the NAHV project, the two values can differ significantly, sometimes by an order of magnitude. In the case study presented in section \ref{results}, the weight is based on \texttt{netEcContribution}, expressed in thousands of euros.

The data is then segmented on a yearly basis. Denoting by $\mathbf{W}_y$ the matrix for a given year $y$, it is a rectangular matrix of size $
       \left| \texttt{projID} \right| 
\times \left| \texttt{orgID}  \right|  
$. The value of each project is assumed to be equally distributed over its duration, from its start date to its end date. Consequently, the contribution of a project to the matrix $\mathbf{W}_y$ is divided proportionally across the years during which the project is active. 
 
Textual fields in table $\mathbf{P}$ contain valuable information, but in a format unsuitable for network analysis. The project objectives are typically described in hundreds of words, and user-defined keywords lack a standardized vocabulary. To make this information usable, the last step in data preparation involves creating Boolean attributes for each project. These attributes encode relevant information in a simplified format, such as whether the project is focused on technology or market uptake, or whether hydrogen is a primary focus or one of several applications. This was accomplished using a script that interacts with the API of a Large Language Model, which processes the lengthy textual fields and generates a dataframe, indexed with \texttt{projID} and one or more Boolean variables, which can be merged with $\mathbf{P}$.

\subsection{Network analysis}
\label{met_network}

The first step in the analysis is to build a network, by identifying its  nodes, edges, and weights in the form of an \textit{adjacency matrix}. 

\begin{figure}[h]
    \centering
    \includegraphics[width=0.5\linewidth]{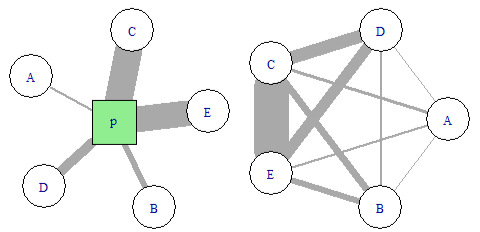}
    \caption{Schematic view of a project as two-mode network (left) and as one-mode network (right). Uppercase letters A, B, C, D, E represent organisations. The green square denoted with P represents a project. Edge width is proportional to the weight, i.e. the value each organisation brings to the project. The sum of weights in the two-mode network is equal to the sum of weights in the one-mode network.}
    \label{fig:network_mode}
\end{figure}

A network (or graph) is a set $G=\{V,E\}$ composed of $n_v = \left| V \right|$ vertices and $n_e = \left| E \right|$ edges. For the purpose of our analysis, a \textit{one-mode network} would be a suitable model, in which nodes represent organisations and edges encode the information about projects. However, in our dataset $\mathbf{W}$ is a rectangular matrix of size $\left|\texttt{projID}\right| \times \left|\texttt{orgID} \right|$, which generates a \textit{two-mode network}, i.e., a network with two distinct sets of nodes where connections are established only between nodes from different sets (e.g., organisations connected through joint participation in projects).
A one-mode network $G$ can be obtained from a two-mode network and described by a square matrix, denoted as $\mathbf{A} = \mathbf{W}^T \mathbf{W}$, where $\mathbf{W}^T$ is the transposed of $\mathbf{W}$. In our case, $\mathbf{A}$ is symmetric, and $G$ is consequently an undirected graph. Figure \ref{fig:network_mode} illustrates the transformation process from the two-mode network of organisations by projects to the one-mode network consisting only of organisation-by-organisation ties.

Since the data is calculated on a yearly basis from 2015 to 2029, the complete model consists of set of networks $\mathcal{G} = \{ G_{2015} \dots G_{2029} \} $ that represents collaborations between organisations within Horizon research project. This spans from the launch of the first projects in 2015 to the anticipated completion of the longest ongoing Horizon Europe project in 2029. The generic network is denoted as $G_y$ where the subscript $y$ refers to a specific year. 

\subsection{Centrality measures analysis}
\label{met-centrality}
The network structure can be evaluated using centrality measures—degree, strength, and coreness—to determine the importance and influence of nodes, each offering distinct insights.

The \textbf{degree} of a node $v$, denoted as $deg(v)$ is the simplest centrality measure, calculated as the number of edges incident to that node. In this context, the degree reflects the number of projects in which an organization is involved during a given year. A higher degree indicates that the node plays a more locally central role within the network, as it has a greater number of direct interactions with other nodes;  in the context of our research, a node with a high degree represents an organisation that is, or has been, a partner in many large projects, reflecting its extensive collaborative involvement.

\textbf{Strength} of a node is the sum of the weights of the edges incident to that node: $s(v) = \sum_{u \in N(v)} w_{vu}$ where $N(v)$ is the set of neighbours  of $v$, and $w_{vu}$ represents the weight of the edge between nodes $v$ and $u$. In weighted networks, strength provides a more nuanced measure of a node's connectivity: with reference to Figure \ref{fig:network_mode}, all nodes have the same degree, but for example $s(C) \gg s(A)$. In this context, the strength reflects the monetary value of the projects in which an organization is involved during a given year.

The k-coreness (or \textbf{coreness}) \cite{Batagelj2003} of a node is a measure of the node's position within the network's hierarchical structure, based on its connectivity. Specifically, a node has a k-coreness of $k$ if it belongs to the k-core of the network. The k-core is a maximal subgraph in which every vertex has at least degree $k$, i.e. within this subgraph, each node is connected to at least $k$ other nodes. In this context, coreness can be interpreted as the capacity of an organisation to partner with other organisations that, in turn, possess the same level of collaborative capacity.

Centrality measures are computed individually for each  $G_{y} \in \mathcal{G}$ and saved as attributes of the nodes in $G_{y}$. This allows to track changes and compare the network structure across different years, providing insights into the dynamics of the collaborations and the shifting roles of organisations over time.

\subsection{Community detection}
\label{met_com_det}
Community detection is a crucial step in the analysis, as it helps to understand the role of an organisation within the network. A \textbf{community} $C$ is defined as a subnetwork of $G$ that satisfies a condition: nodes that belong to $C$ are more densely connected within each other than with the rest of $G$. A \textbf{partition} $\mathcal{P}$ is a set of $k$ disjoint subnetworks $C_1, \ldots, C_k$ whose union is equal to $G$.

A community detection algorithm $\mathcal{A}(G, \rho) \rightarrow \mathcal{P}$ is a function that takes as input a graph $G$ and one or more parameters $\rho$, and returns a partition $\mathcal{P}$. Several community detection algorithms are discussed in literature \cite{Diboune2024,Khawaja2024}. Ideally, any algorithm should produce a single, valid partition each time it is applied with the same parameters. In practice, however, for large, dense networks, several issues may arise. Some community detection algorithms may produce \textbf{invalid results} such as communities that are internally disconnected or fail to meet the basic criterion of being more densely connected within than with the rest of the network. This issue should be addressed with a thorough quality check on $\mathcal{P}$.
A less known issue is the \textbf{Input-Ordering Bias}: Although networks are mathematically non-ordered entities, their implementation in any software algorithm is inevitably ordered (in the form of a list of edges, or a matrix as $\mathcal{A}$ mentioned above). The order in which nodes and edges are stored in the practical implementation of the network can affect the results, as illustrated in \cite{morea2024enhancingstabilityassessinguncertainty}. 

A more elusive issue is the \textit{variability} of results. When $\mathcal{A}$ is based on a heuristic or randomized approach to identify the optimal partition, it may yield different results each time they are executed: $\mathcal{P}_i \ne \mathcal{P}_j$.   All these issues can be addressed as per Algorithm 2, based on the repeated execution of $\mathcal{A}(G^*,\rho)$, where $G^*$ is shuffled version of $G_{y}$ to reduce the effect of input ordering bias) and to explore the \textit{solution space}. Specifically, the \textbf{solution space} $\mathbb{S} = \{ \mathcal{P}_1, \mathcal{P}_2, \ldots, \mathcal{P}_{ns} \}$ is the set of all unique partitions that $\mathcal{A}$ produces across $t$ trials. Our research aims to analyse $\mathbb{S}$, and to determine the minimum number of trials $t_c$ required to confidently assert the completeness of $\mathbb{S}$.

After selecting the most suitable algorithm, the analysis may yield either a single, ideal solution or a dominant solution—one that is more frequently observed than others. In these cases, this solution can be identified and analysed. However, if no single solution emerges as dominant, a "consensus community detection approach" can be applied to consolidate the variable results into a single, reliable solution. This process is illustrated in Algorithm 2 and discussed thoroughly in  \cite{morea2024enhancingstabilityassessinguncertainty}.

\begin{algorithm}
\caption{Consensus Community Detection}
\begin{algorithmic}[1]
    \For{$t = 1$ to $t_{max}$} 
        \State $G^* \gets$ random permutation of $G_{y}$
        \State $\mathcal{P}_t \gets \mathcal{A}(G^*, \rho)$ 
        \If{$\mathcal{P}_t$ is identical to another partition in $\mathbb{S}$} 
            \State Update frequency count in $\mathbb{S}$
        \Else
            \State Append $\mathcal{P}_t$ to $\mathbb{S}$
        \EndIf
    \EndFor
    \State \textbf{Identify a single solution from the solution space}
    \If{a single or dominant solution exists} 
        \State $\mathcal{P}_{y} \gets \mathcal{P}_t$
    \Else
        \State $\mathcal{P}_{y} \gets \text{CCD}(\mathbb{S})$ \Comment{Apply consensus community detection on $\mathbb{S}$}
    \EndIf
\end{algorithmic}
\end{algorithm}

\begin{figure}[h]
     \centering
     \includegraphics[width=0.95\linewidth, height = 4cm]{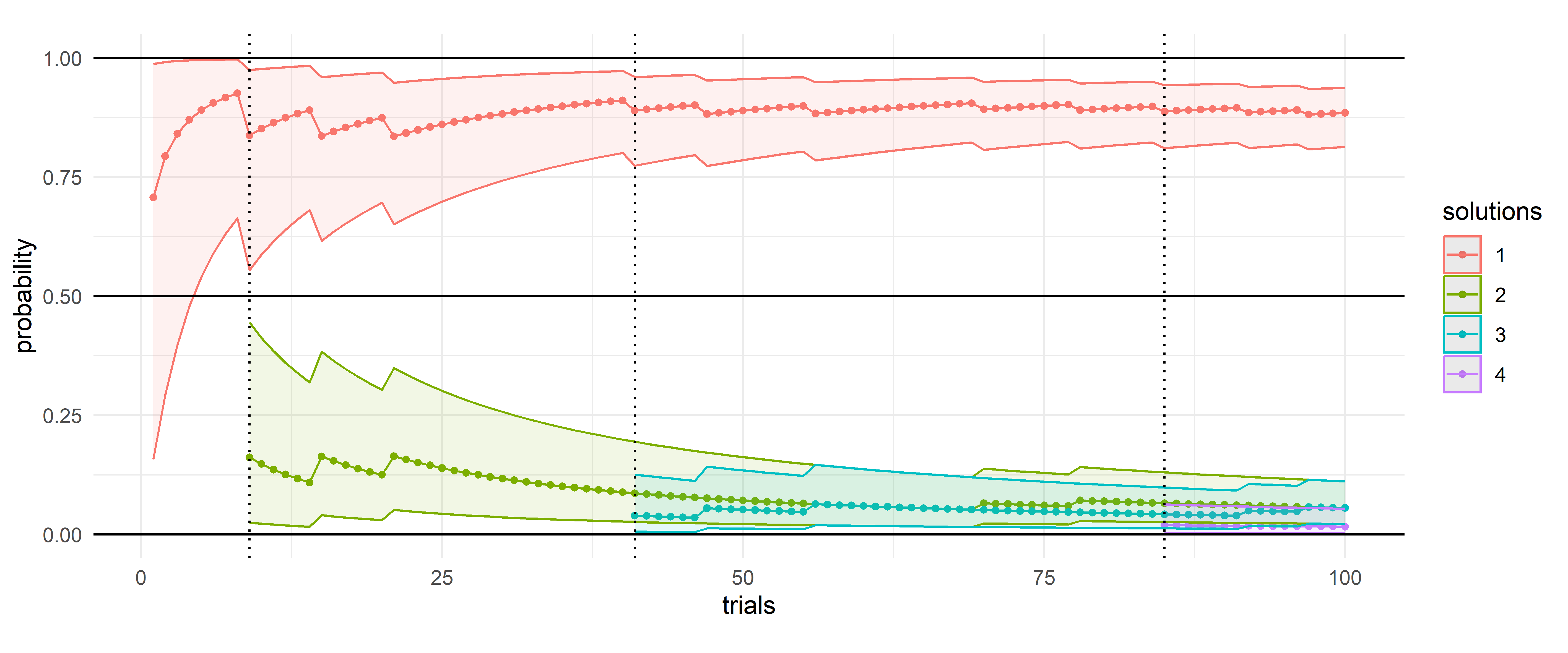}
     \caption{Confidence intervals associated with different solutions in the solution space.}
     \label{fig:confidence}
 \end{figure}

Knowledge about the solution space, along with the awareness that a single or dominant solution exists, emerges gradually as a result of an experiment of repeated trials $\mathcal{A}(G, \rho) \rightarrow \mathcal{P}_i$, as illustrated in Figure \ref{fig:confidence}. In the example, the probability of solutions in $\mathbb{S}$ are represented as point estimates (solid line for $\bar{p}$) and intervals (ribbon between $(p_{lower}$ and $p_{upper}$). A prevalent solution (shown in red) emerges soon; as $t$ increases additional solutions are discovered and beyond $t=50$, the probability distributions are virtually unaffected by any new solutions. When the experiment is repeated, a slightly different situation may be observed, but convergence towards a dominant solution consistently occurs.

Probabilities associated with each solution is $\mathbb{S}$ calculated under a Bayesian framework, using a Beta Binomial model: this ensures a thorough yet efficient analysis, and optimizes computational resources. 

\subsection{Temporal evolution of communities}
\label{met_tempo}
As introduced in Section \ref{intro}, it is important to provide  insights into the underlying dynamics of collaboration over time, which can be achieved by tracking and categorize the evolution of communities over time. The information  is represented as a family of networks, denoted as $G_{y}$ where each network corresponds to a specific year. Within each yearly network, communities are identified, and our primary objective is to understand how these communities evolve over time. The method involves comparing each community $Ci \in G_{y}$ with each $Cj \in G_{y+1}$. Through this comparison, we determine whether $C_i$ and $C_j$ are disjoint or have a non-null intersection. If they intersect, the relationship between the two communities is further classified as either continuing (i.e. $C_i$ shares most of its members with $C_j$), or as part of a merge or split. To ensure accurate tracking across different years, we assign global community labels that remain consistent for identical or continuing communities.

\section{ Results}
\label{results}
This section applies the methodology outlined in Section \ref{met} to a subset of hydrogen-related projects, including the NAHV project, which is the focus of this study; the data encompasses projects from Horizon 2020 and Horizon Europe, covering the period from 2015 to 2029. The analysis aims to investigate leadership roles, community structures, and their evolution over time. Furthermore, the analysis demonstrates how AI-generated keywords can be integrated to provide additional insights into the distinction between market-oriented and technology-oriented projects.

\begin{figure}[H]
    \centering
    \includegraphics[width=.9\linewidth]{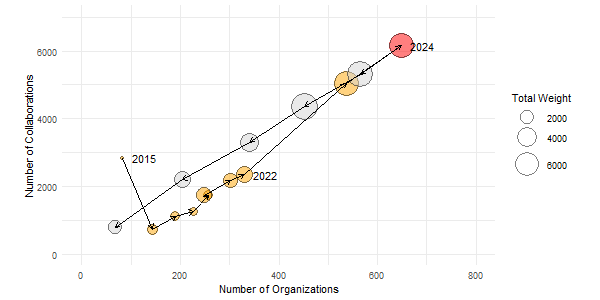}
    \caption{Size of the networks $G_{y}$ represented by the number of organisations (horizontal axis) and the number of collaborations (vertical axis) and total investment per year (bubble size). Between 2016 and 2024 there is an increase in the number of organisations from 83 to 648 and a proportional rise in collaborations as represented by weights, rising from 547 to 6628. Current data for projects continuing in years 2025 to 2029 suggest further growth in the total value of projects. }
    \label{fig:networksyears}
\end{figure}

The data has been prepared using the \textit{EuroSciVoc} topic \textit{hydrogen energy}, and the AI-generated categories are market vs technology, as explained in Section \ref{met_data_prep}. The family of networks $\{ G_{2015} \dots G_{2024} \}$ has a remarkable evolution over time. $G_{2015}$ is composed of 83 organisations, and the number steadily increases to 648 by 2024. The number of collaborations, represented by the network's edges, grew proportionally during this period, and the total value of projects, measured by the sum of weights in the network, saw an even more significant increase, rising from 547 to 6628. 
Looking ahead to the period from 2025 to 2029, the future of collaborations will be shaped by contracts currently in operation and influenced by upcoming calls and new projects. As shown in Figure \ref{fig:networksyears}, after 2024 the number of partners and collaborations decreases, while the total investment remains stable for the next two years. 

\subsection{Leadership roles}

\begin{figure}[h]
    \centering
    \includegraphics[width=0.95\linewidth]{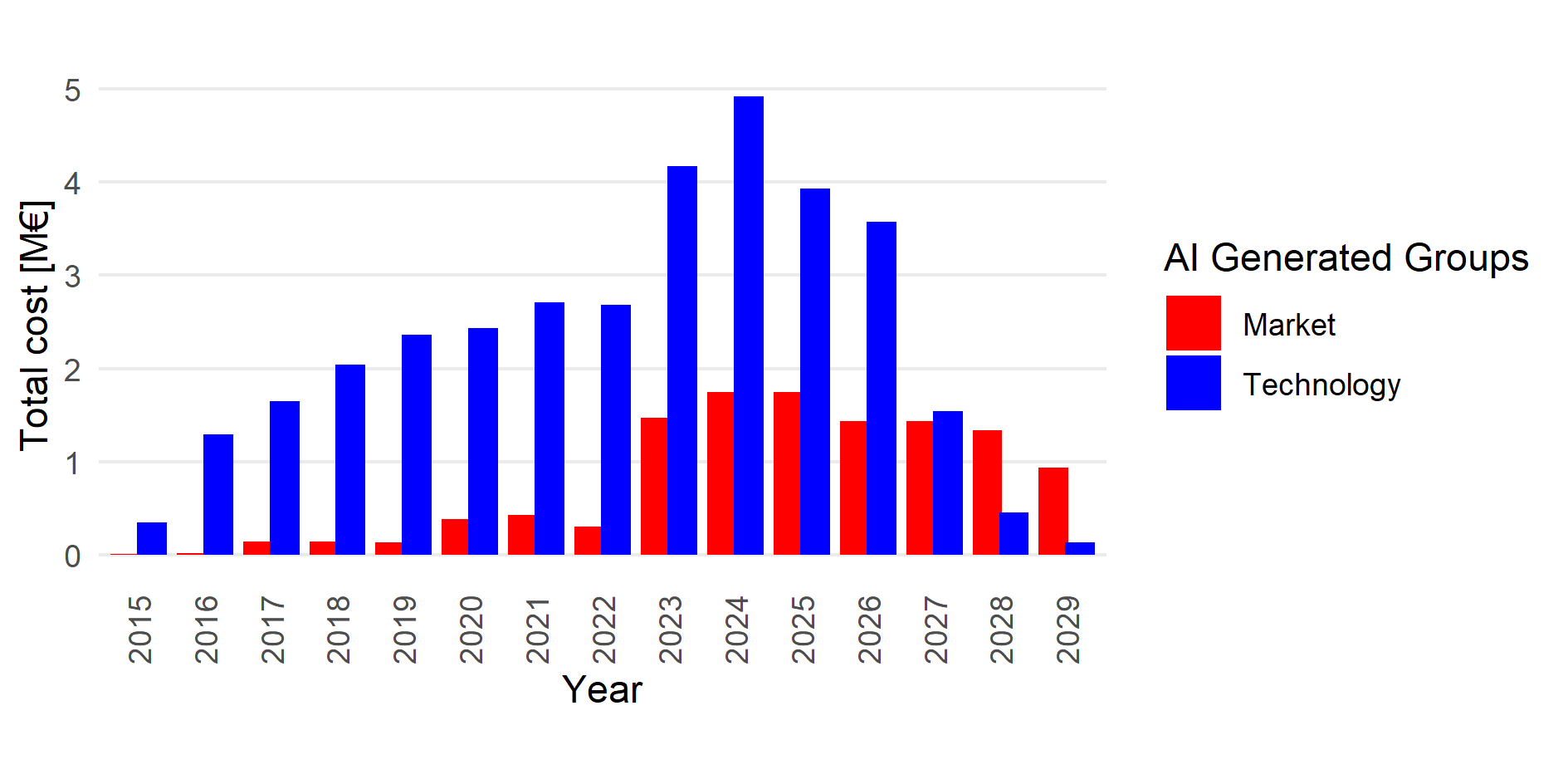}
    \caption{Comparison between technology-oriented and market-oriented project. The value is measured by \texttt{netEcContribution}; groups are identified by the AI generated labels: $G^{M}_{y}$ and $G^{T}_{y}$}
    \label{fig:AIgengroups}
\end{figure}

The methodology for AI-generated categories, discussed in \ref{met_data_prep}, is applied to the hydrogen projects dataset. This classification approach distinguishes between projects focused on market development — encompassing policy, market uptake, business models, and hydrogen valleys — and those centred on technological development. The NAHV project clearly falls into the former category. This classification enables the creation of two distinct families of networks, denoted as $G^{M}_{y}$ and $G^{T}_{y}$. The total value of the projects in each group is shown in Figure \ref{fig:AIgengroups}, revealing an interesting trend: while investment in market-oriented projects has been consistently present since 2016, there is a notable increase in such investments starting in 2023.
 
\begin{figure}[H]
        \includegraphics[width=7cm, height=6cm]{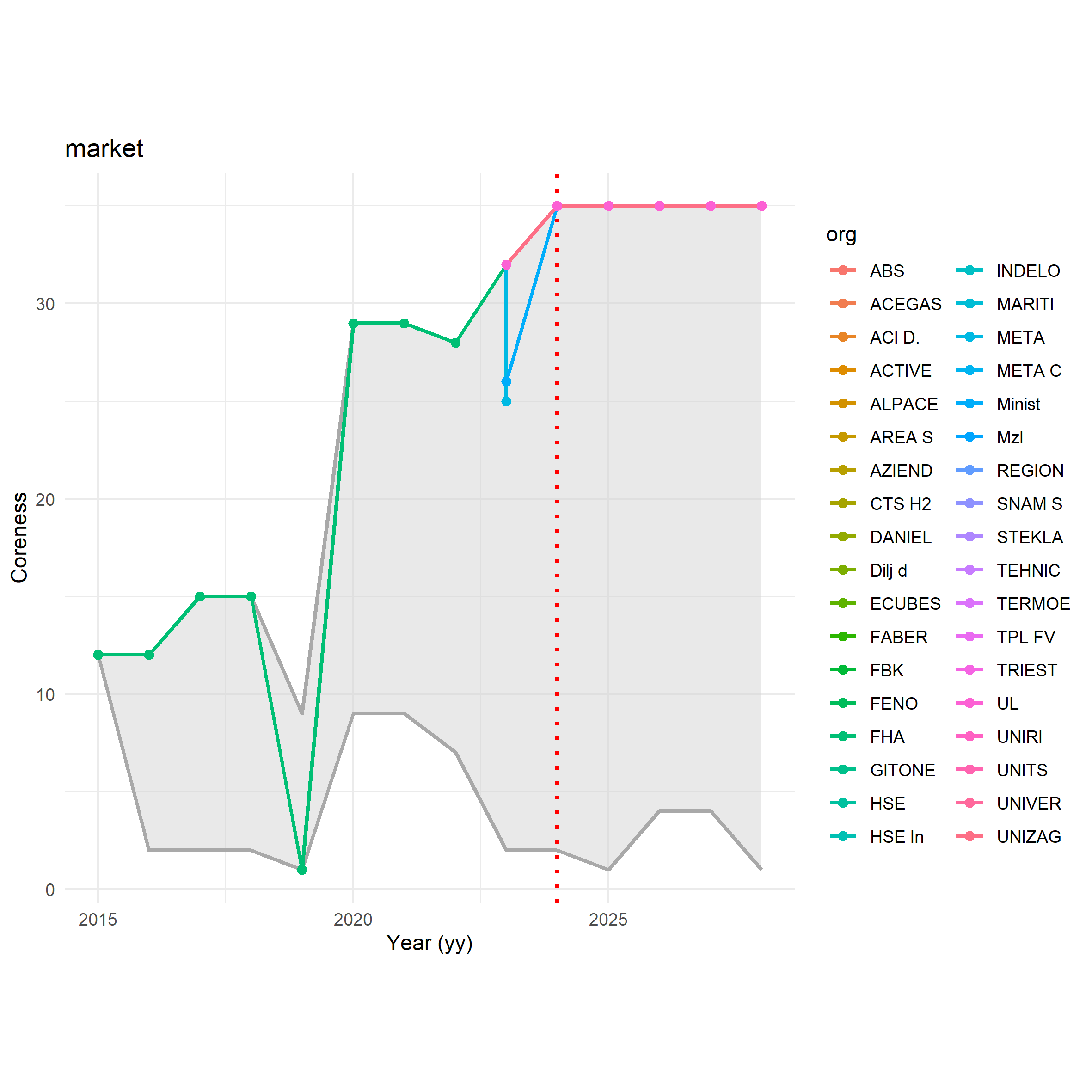}
        \includegraphics[width=7cm, height=6cm]{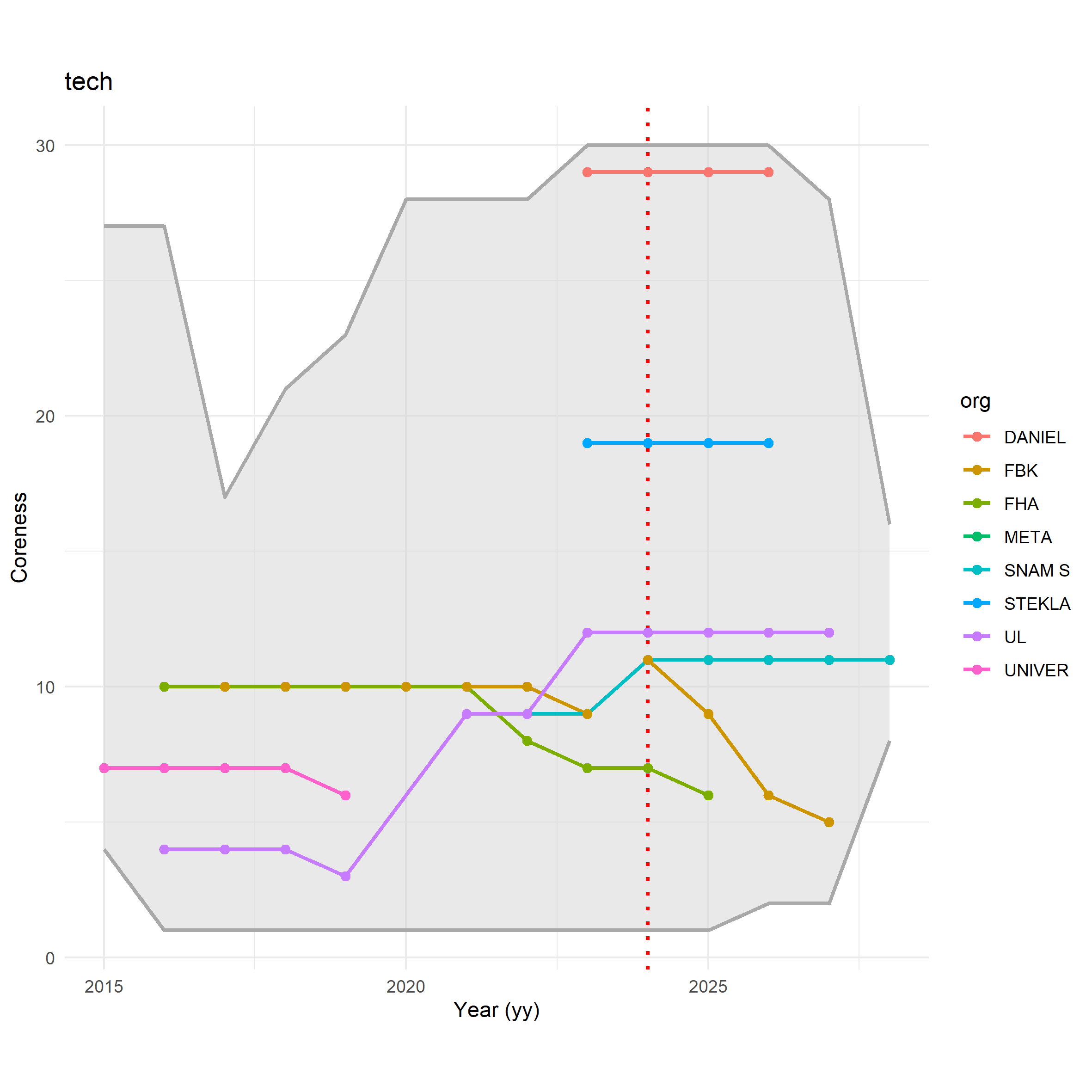}
    \caption{Coreness of organisations involved in the NAHV project, shown in comparison with the range of coreness values for each year (grey background).}
    \label{fig:coreness}
\end{figure}

 Figure \ref{fig:coreness} illustrates how centrality measures and AI-generated categories provide insights into the roles of organisations in hydrogen-related innovations, highlighting differences between those engaged in market-oriented and technology-oriented projects.

The figure shows in grey the range of values found each year, and highlights organisations participating in the NAHV project. The diagrams provide an overview of the evolution of coreness in $G^{M}_{y}$ (left) and $G^{T}_{y}$ (right). The grey shaded area represents the range of coreness values observed each year. The organisations involved in the NAHV project are highlighted with coloured dots, connected by lines that illustrate their progression over time. A notable observation is that, while all NAHV partners are part of $G^{M}_{y}$ (which is expected, given NAHV's classification as a market project), only a few are involved in $G^{T}_{y}$.

Coreness values can fluctuate over time: an increase suggests that an organisation is gaining influence through participation in relevant projects, while a decrease may indicate that, following the completion of a project, the organisation has not yet initiated a new one within the Horizon framework. The analysis also includes the years 2024 to 2029, reflecting ongoing projects that are scheduled to conclude during this period. It is important to note that this is not a forecast, but rather a record of existing projects with future completion dates.

Moreover, centrality of all organisations can be visualized in a degree-coreness plane, where each organisation is represented by a bubble. Degree serves as a proxy for an organisation's capacity to form partnerships with many other organisations, while coreness reflects its ability to partner with influential organisations. Although coreness is limited by degree, the size of the bubble represents the organisation's strength, which serves as a proxy for its capacity to attract substantial funding and invest in large projects. An example is shown in figure \ref{fig:degree-centrality}, which also includes an information about communities.

\subsection{Communities}
In the context of this paper, communities refer to relevant subgroups of organisations that exhibit strong collaborative ties. The focus is on analysing a single partition and its evolution over time. The analysis is conducted on the family of complete networks, $G_{y}$.

\begin{figure}[H]
    \centering
    \includegraphics[width=0.85\linewidth]{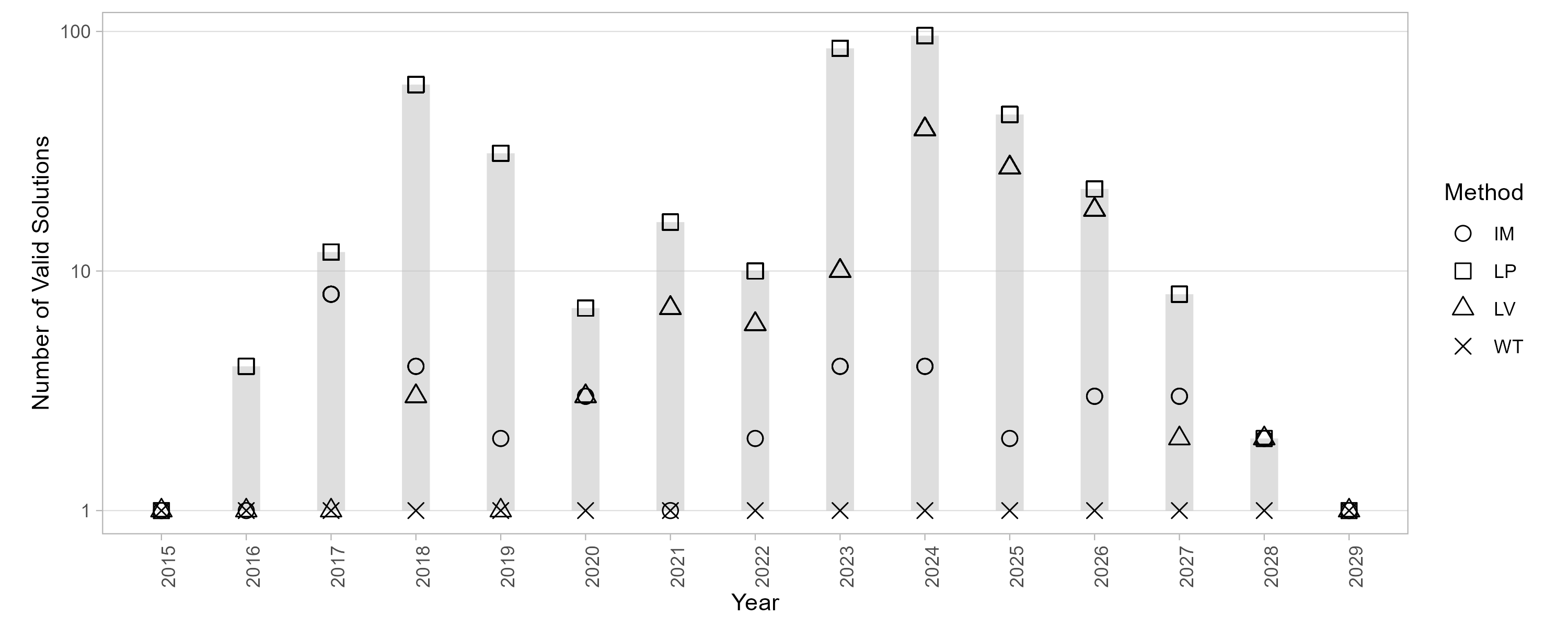}
    \caption{Number of solutions $\left| \mathbb{S}_y \right|$ generated by different community detection algorithms.}
    \label{fig:n-solutions}
\end{figure}

As explained in Section \ref{met_com_det}, the first step involves exploring the solution space $\mathbb{S}^A_y$ in terms of completeness, the number of solutions, and their validity across various algorithms. The tested algorithms include Louvain (LV) \cite{blondel2008Louvain}, Leiden (LD) \cite{traag2019leiden}, Label Propagation (LP) \cite{raghavan2007labelprop}, Edge Betweenness (EB) \cite{newman2004edgebetw}, Leading Eigenvector (EV) \cite{newman2006EV}, Walktrap (WT) \cite{pons2005walktrap}, and Infomap (IM) \cite{rosvall2007infomap}. The tests were conducted in R \cite{R-base}, using the iGraph \cite{igraph} and communities \cite{communities_package} libraries.

The number of solutions $\left| \mathbb{S}^A_y \right|$ are illustrated in \ref{fig:n-solutions}: most algorithms produced multiple solutions across most years, with the exceptions of 2015 and 2029, where the networks were simpler. Interestingly, WT consistently provided a unique solution. However, EB had a different issue, generating partitions that were not similar to each other, with similarity coefficients often below 0.5. EV produced multiple valid solutions in most years but failed to find a partition in 2024.

\begin{figure}[h]
    \centering
    \includegraphics[width=0.85\linewidth]{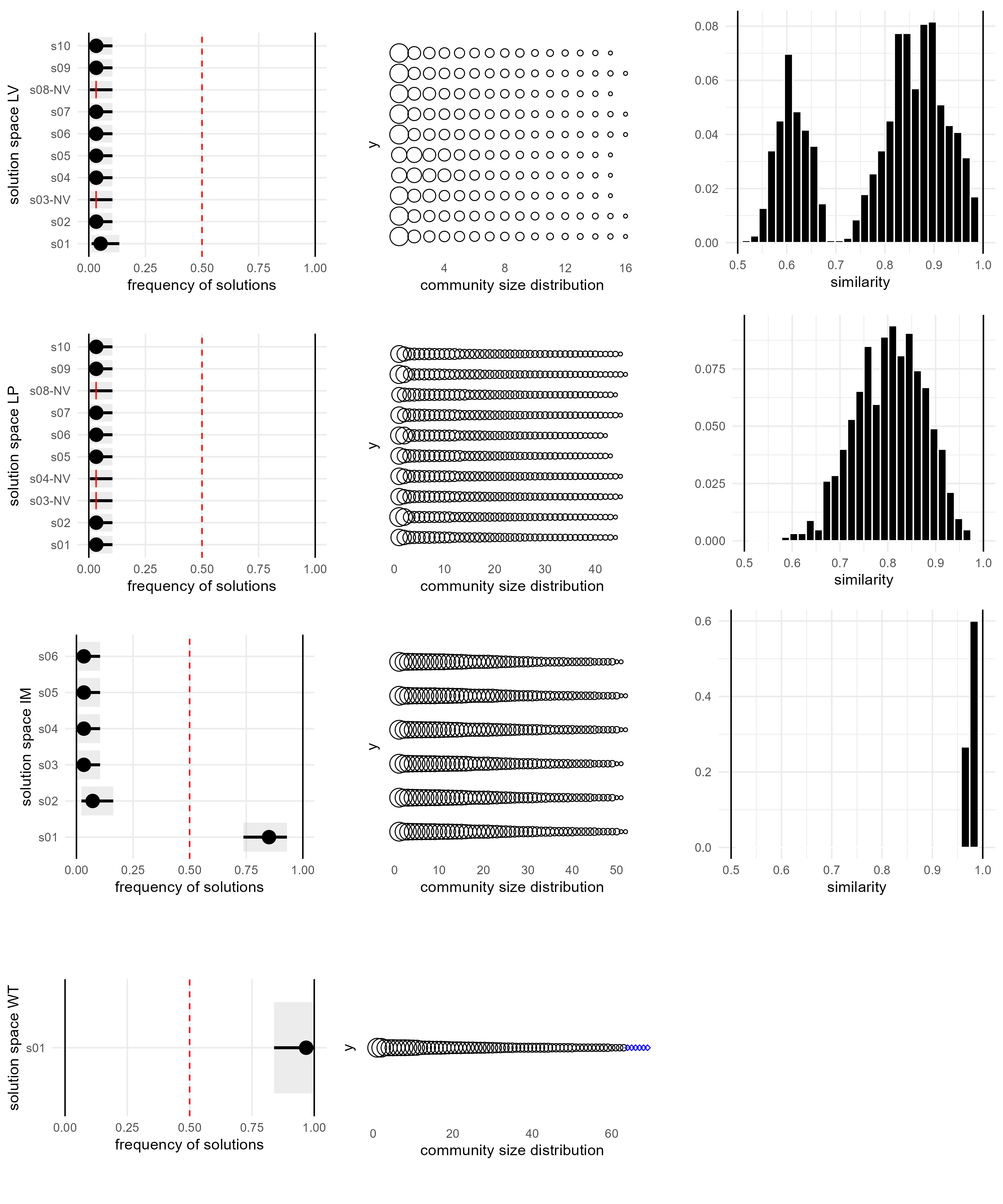}
    \caption{Solution space diagrams. Left: frequency and confidence intervals for the solutions identified by each algorithm. In the case of LV and LP only the 10 most frequent solutions are shown. Middle: distribution of community size for each solution. Proper communities are represented as black circles, single-node communities are represented as blue diamonds. Right: pairwise similarity between solutions.}
    \label{fig:sol-space}
\end{figure}

A more in-depth observation of the solution space for 2024 (the most critical year) reveals additional details, as shown in \ref{fig:sol-space}. LV generates a large set of solutions with $| \mathbb{S}^{LV}_{2024} | = 69$, which are also highly diverse, as evidenced by the similarity plot showing a bimodal distribution with peaks around 0.6 and 0.9. Moreover, approximately one-third of the solutions are invalid due to containing one or more internally disconnected communities. This is a well-documented issue of the algorithm, as discussed in the literature. LD, specifically designed to address the problem of disconnected communities, also produced invalid solutions for this network: its partition consists of a few large communities and a high proportion of singletons, with a mixing parameter exceeding 0.5, indicating that nodes within a community were more connected to nodes outside their community than to those within it.

LP generates a different solution with each run; in this case, with $t = 100$, there are 96 valid and distinct solutions, while 4 are invalid due to internally disconnected communities. Better results can be achieved with IM that produces  $| \mathbb{S}^{IM} | = 1$ in most years, and a dominant one solution in the most complex cases such as 2024. Overall, WT is the only algorithm that consistently provided valid results across all years, with valid partitions and $| \mathbb{S}^{WT} | = 1$, consequently, it will be used for community detection and temporal evolution analysis.
 
\begin{figure}
    \centering
    \includegraphics[width=0.85\linewidth]{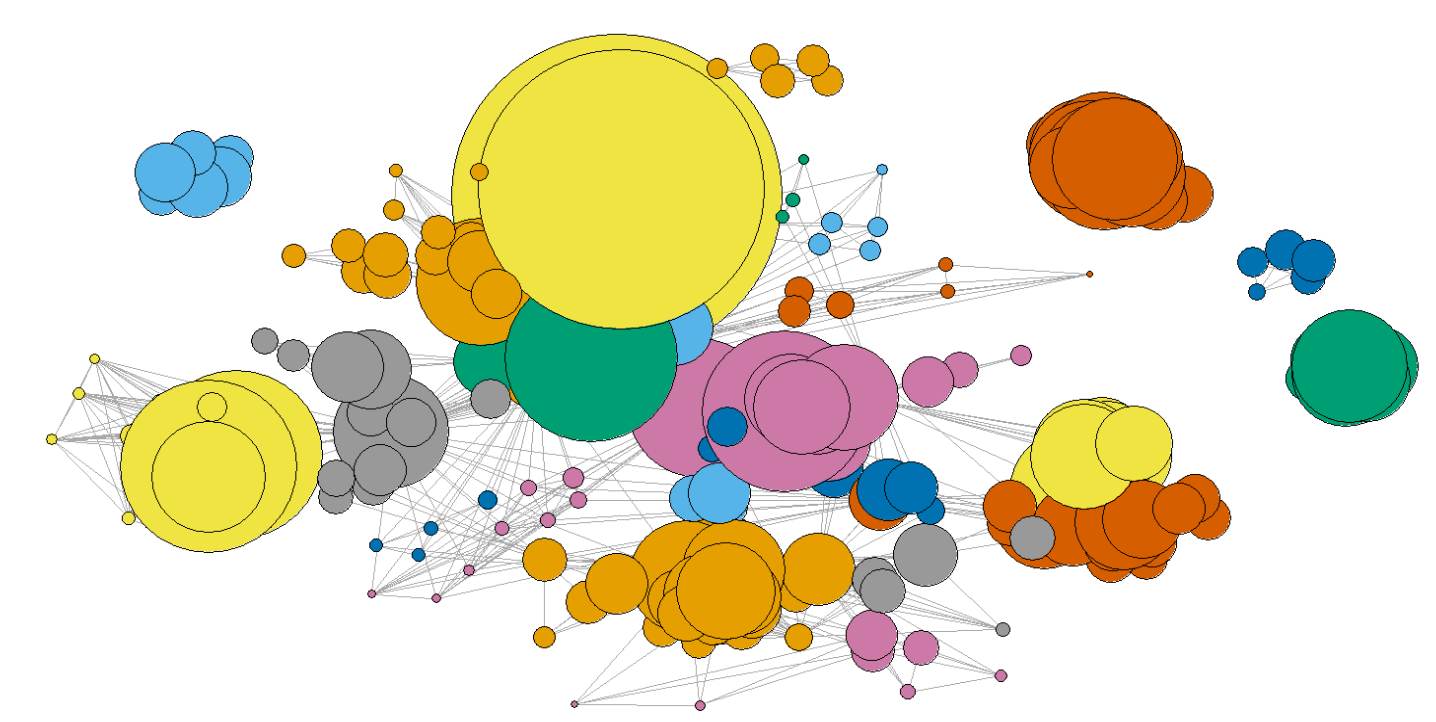}
    \caption{Network $G_{2018}$, with communities grouped and highlighted in different colours.}
    \label{fig:commnetwork2018}
\end{figure}

The selected partition $\mathcal{P}_y$ can be represented as a network in which nodes belonging to the same community are grouped and coloured to distinguish between communities. For example, the result for $P_{2018}$ is shown in Figure \ref{fig:commnetwork2018}. Some communities appear disconnected from the others; these consist of organisations that, in that year, participated in only one project and thus form a separate component.

Community detection can be integrated with centrality measures to provide a more comprehensive understanding of network structure. 

\begin{figure} [H]
    \centering
    \includegraphics[width=0.85\linewidth]{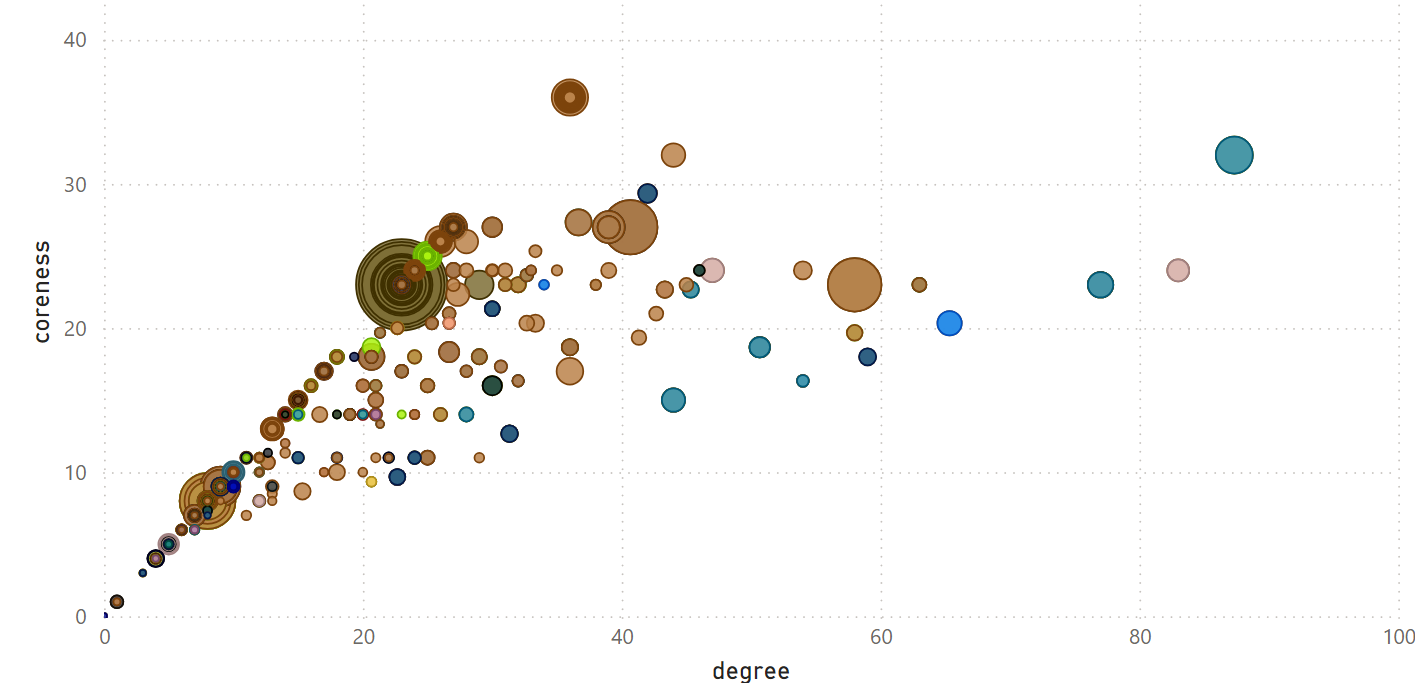}
    \caption{Centrality measures of all organisations in $G_{2024}$. Each bubble represents an organisation in the degree-coreness plane. Degree is a proxy for the organisation's ability to win projects and attract funding, while coreness reflects its capacity to partner with other influential organisations. The largest bubbles in the top-right of the diagram highlight organisations that excel in both areas.}
    \label{fig:degree-centrality}
\end{figure}

Figure \ref{fig:degree-centrality} illustrates the results for $\mathcal{P}_{2018}$. In this visualisation, each bubble represents an organisation, with its position determined by coreness and degree centrality. Bubble colours correspond to the community, and  bubble size is proportional to the strength (the total weight of its connections in the given year). The diagram clearly demonstrates that degree is the upper limit for coreness, as no points appear above the line with a slope of 1. In practical terms, this indicates that an organisation that participates in many projects (high degree) is not necessarily the most influential in the network. Strength, which reflects the total weight of an organisation's connections, remains independent of both degree and coreness, meaning that large bubbles can appear anywhere on the graph. This means that some organisations, despite having relatively few connections (low degree), are linked to relevant partners (high coreness) and stand out for their ability to secure significant EU research funding (high strength).

\subsection{Evolution of communities}
\begin{figure}[h]
    \centering
    \includegraphics[width=.95\linewidth]{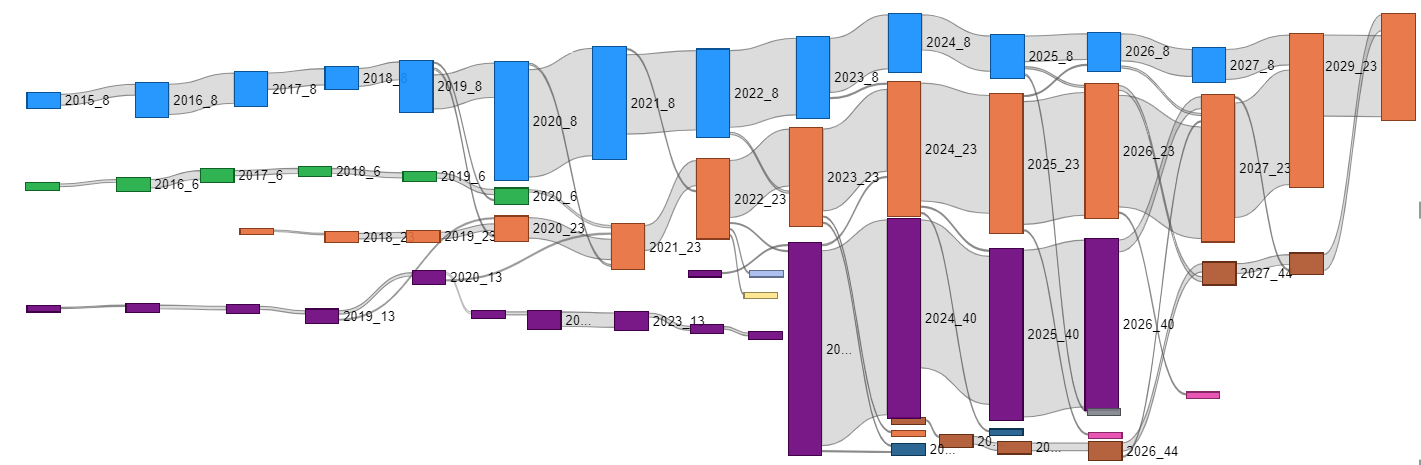}
    \caption{Evolution of the largest communities from 2015 to 2029. For clarity, only the six largest communities are displayed.}
    \label{fig:tempo}
\end{figure}

Temporal evolution was calculated as described in Section \ref{met_tempo}, using the WT algorithm to ensure the robustness of the results.  In the diagram shown in  Figure \ref{fig:tempo}, the horizontal axis represents time, with each community depicted as a coloured rectangle. The height of each rectangle is proportional to the sum of the strengths of its members. For the purpose of clarity, only the six largest communities are represented. 
The analysis reveals a significant pattern: initially a large number of small communities is formed, and their configuration changes rapidly. This reflects the fact that in Horizon 2020 public funding was primarily allocated to technology-oriented research projects, and partnerships changed rapidly. However, with the onset of Horizon Europe, and specifically the Clean Hydrogen partnership, three large communities emerge, and are set to continued collaboration through 2029. 

The significant size of the largest communities reflects the increasing allocation of resources to Horizon Europe projects in the hydrogen sector. This growth is driven by the development of hydrogen valleys, which have fostered long-term, stable partnerships, contributing to both the stability of collaborations and the continued expansion of these communities.

\section{Discussion}
This paper presented a comprehensive methodology for analysing the long-term impact of EU-funded research projects, with a case study focused on the hydrogen sector. By employing network analysis, centrality measures, and community detection, we demonstrated how to identify pivotal organizations, monitor their influence over time, and delineate the evolution of collaborations fostered by Horizon 2020 and Horizon Europe projects.

Specifically, the analysis shows that Horizon grants in hydrogen sector projects have prompted the emergence of communities that have significantly evolved over the years. Initially characterized by small, rapidly changing groups, these groups gradually shifted towards the formation of larger, more stable partnerships, particularly in the domain of hydrogen valleys.

The methodology presented in this paper, particularly the use of centrality measures, community detection algorithms and temporal network analysis, offers opportunities to improve the visualizations in the Horizon Dashboard \cite{HorizonDashboard}. For instance, the "Organization Profile" visualization, which allows users to explore an organization's performance in EU research and innovation programs, could benefit from incorporating centrality measures and tracking community memberships of each organization over time. This could be done by analyzing the entire Horizon network or by focusing on specific subsets categorized according to the top-level EuroSciVoc fields: \textit{Natural Sciences}, \textit{Engineering and Technology}, \textit{Medical and Health Sciences}, \textit{Agricultural Sciences}, \textit{Social Sciences}, and \textit{Humanities}. 
 
 One limitation of this study is that it is focused exclusively on the CORDIS dataset, which restricts its scope to EU-funded initiatives, potentially neglecting initiatives funded by national or local authorities, or private funds. Incorporating data from other sources could provide a more comprehensive perspective, but such datasets are currently unavailable.

\section{Conclusions and future work}
The methodology presented in this paper can provide valuable insights for both policymakers and project coordinators. Policy makers, such as those supporting Hydrogen Valley projects across Friuli Venezia Giulia, Croatia, and Slovenia, can leverage this methodology to monitor the engagement of organizations in their respective territories. As the Hydrogen Valley projects launched in 2022-2023 come to an end and new Horizon calls for proposals become available, policy makers will be able to assess whether today's leading organisations can form new partnerships and maintain their leadership positions. 
In addition, policy makers at European level can use the methodology to gain insights into the evolution of communities, to assess their openness to newcomers and the stability of collaboration between industrial and academic partners.
From an alternative perspective, project coordinators within industrial or research organisations may use the methodology presented in this paper to identify potential partners who may be influential in facilitating future collaborations. 

Future work will focus on several key areas. We will continue to monitor the centrality measures of the NAHV partners over the course of the project and beyond. In addition, we will apply the methodology to further case studies on different topics, in particular pandemic preparedness and electron microscopy, both of which are topics of interest for Area Science Park and have been published in \cite{horizon_projects_network_2024}. Finally, future research will explore how the rich textual information from project deliverables available in CORDIS can be effectively leveraged to enhance network models.

\section*{Data availability}
The underlying data for this research is openly available through  data.europa.eu, the official portal for European data. Specifically, data about Horizon 2020 projects are available from: \href{https://data.europa.eu/data/datasets/cordish2020projects}{https://data.europa.eu/data/datasets/cordish2020projects} and data about Horizon Europe projects are available from: \href{https://data.europa.eu/data/datasets/cordis-eu-research-projects-under-horizon-europe-2021-2027}{https://data.europa.eu/data/datasets/cordis-eu-research-projects-under-horizon-europe-2021-2027}. 

Data generated by this research is published in Zenodo, under the terms of the Creative Commons Attribution 4.0 International license (CC-BY 4.0), as 'Horizon Projects Network':

\href{ https://doi.org/10.5281/zenodo.13765372}{https://doi.org/10.5281/zenodo.13765372}  

\section*{Software availability}
Source code available from: \href{https://github.com/fabio-morea/horizon-intelligence}{https://github.com/fabio-morea/horizon-intelligence}

\section*{Author contributions}
F.M. conceptualized the study, developed the code, prepared the data, conducted the analysis, and drafted the manuscript. A.S. contributed to defining the analytical framework on hydrogen valleys, analysis, and conclusions. D.D.S. contributed to the methodology and supervised the analysis. All authors contributed to refining the manuscript. 

\section*{Acknowledgments}

Authors F.M. and A.S. acknowledge the financial support of the North Adriatic Hydrogen Valley (NAHV) project. The project received support from the European Union under grant agreement 101111927. The funders had no role in study design, data collection and analysis, decision to publish, or preparation of the manuscript.
The views and opinions expressed in this paper are solely those of the author(s) and do not necessarily reflect the official positions of the European Union or the Clean Hydrogen Joint Undertaking. Neither the European Union nor the granting authority assumes responsibility for the content presented.

\bibliographystyle{ieeetr}

\bibliography{references}

\end{document}